\documentclass[conference]{IEEEtran}
\usepackage{times}

\usepackage[numbers]{natbib}
\usepackage{multicol}
\usepackage{graphicx}
\usepackage{subcaption}
\usepackage{wrapfig}
\usepackage[bookmarks=true]{hyperref}

\begin{document}

\title{Evaluating a VR System for Collecting Safety-Critical Vehicle-Pedestrian Interactions}

% You will get a Paper-ID when submitting a pdf file to the conference system
% \author{Author Names Omitted for Anonymous Review.}

\author{
\authorblockN{Erica Weng}, Kenta Mukoya, Deva Ramanan, and Kris Kitani
\and
\authorblockA{Robotics Institute\\
Carnegie Mellon University\\
Pittsburgh, PA\\
}
}

% avoiding spaces at the end of the author lines is not a problem with
% conference papers because we don't use \thanks or \IEEEmembership

\author{\authorblockN{Erica Weng\authorrefmark{1}\authorrefmark{3},
Kenta Mukoya\authorrefmark{2},
Deva Ramanan\authorrefmark{1}, and
Kris Kitani\authorrefmark{1}}
\authorblockA{\authorrefmark{1}Robotics Institute\\
Carnegie Mellon University\\
Pittsburgh, PA 15217}
\authorblockA{\authorrefmark{2}Denso Japan. Work done while visiting at the Carnegie Mellon University Robotics Institute.\\
}
\authorblockA{\authorrefmark{3}Correspondence to: eweng@cmu.edu}
}

\maketitle

\begin{abstract}
  Autonomous vehicles (AVs) require comprehensive and reliable pedestrian trajectory data to ensure safe operation. However, obtaining data of safety-critical scenarios such as jaywalking and near-collisions, or uncommon agents such as children, disabled pedestrians, and vulnerable road users poses logistical and ethical challenges. This paper evaluates a Virtual Reality (VR) system designed to collect pedestrian trajectory and body pose data in a controlled, low-risk environment. We substantiate the usefulness of such a system through semi-structured interviews with professionals in the AV field, and validate the effectiveness of the system through two empirical studies: a first-person user evaluation involving 62 participants, and a third-person evaluative survey involving 290 respondents. Our findings demonstrate that the VR-based data collection system elicits realistic responses for capturing pedestrian data in safety-critical or uncommon vehicle-pedestrian interaction scenarios. 
\end{abstract}

\IEEEpeerreviewmaketitle

\begin{figure}[ht]
  \includegraphics[width=\columnwidth]{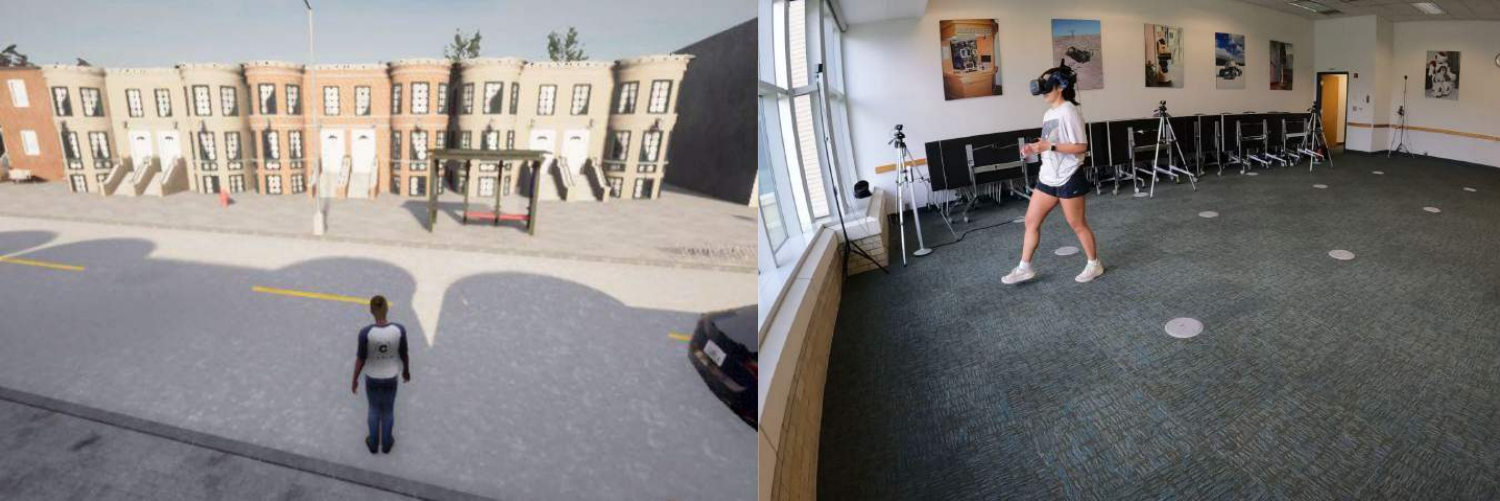}
  \caption{Left: Participant's avatar within the virtual environment from a third-person perspective. Right: participant wearing the VR headset in the data collection environment.}
  \label{fig:teaser}
\end{figure}

\section{Introduction}
Pedestrian trajectory data collection is necessary for training safe and robust pedestrian forecasting models for autonomous vehicles (AVs),
but many important vehicle-pedestrian interaction scenarios are missing in existing popular trajectory and tracking datasets~\cite{Caesar2019-az,Ettinger2021-yl,Geiger2013-xv,Chang2019-jr,Wilson2023-yc}.
Specifically, collecting data in many scenarios is difficult.
% such as scenes that feature close interactions between vehicles and pedestrians, vulnerable road users such as people with disabilities, or children.
% Furthermore, many of these scenarios are difficult or impossible to collect data for in real life. 
For example, it is difficult to collect data of jaywalkers:
pedestrians who cross a street at a location other than designated crosswalks or intersections, often in disregard of traffic regulations. 
Jaywalking is generally considered unsafe and is illegal in many regions, 
and thus it is sparsely observed in real life or real datasets. 
However, it is still important that forecasting models understand the behavior of jaywalkers for pedestrian safety.
Furthermore, it is problematic to willfully direct pedestrians to perform illegal behavior for the purpose of data collection. As another example, real-life datasets have a deficiency of data of children pedestrians~\cite{Caesar2019-az,Ettinger2021-yl,Geiger2013-xv,Chang2019-jr,Wilson2023-yc}. Child street-crossing behavior differs from that of adults; children often take more risks and appear more unpredictable~\cite{Gitelman2019-hc}. Due to child protection laws, it is even more problematic to put children in potentially dangerous situations for the purpose of data collection~\cite{Office_for_Human_Research_Protections_OHRP2010-di}. As a final example, there exist very specific vehicle-pedestrian interaction scenarios,
such as walking on the road alongside traffic, crossing a parking lot or garage entrance with vehicles entering and exiting, 
or crossing a narrow urban alleyway in which the building structure obstruct easy view of oncoming vehicles.
These scenarios arise due to specific urban structural and geographic features that may only be sparsely found in real-life urban areas. 
% as sites for targetted data collection. present in many places, 
However, these scenarios are still relevant and important to robust pedestrian safety,
and thus it is important that data is collected for these scenarios.
% DEVA: I think that narratively, it feels natural to point to point out the body of part on 
% ERICA: I added a section in the intro for simulated synthetic data, as well as related works.
There is a large body of work on controllable simulators that can generate large quantities of synthetic data
of various scenarios of choice~\cite{Dosovitskiy2017-oy,Chen2021-yg,Yang2023-dk},
including the aforementioned scenarios that are uncommon in real-life but critical to pedestrian safety.
There is also much work done on AV perception, prediction, and planning models that utilizes simulated data for training~\cite{johnson2016driving, huang2017expecting}. 
Relevant to our area of interest, \citet{Huang2017-vy} utilize synthetic pedestrian data to train a detector to detect unusual or uncommon pedestrian behaviors.
However, all such methods that train largely on synthetic data suffer from the well-known "sim2real" gap.
This gap exists because simulators struggle to reproduce certain aspects of reality: 
for example, in CARLA~\cite{Dosovitskiy2017-oy}, vehicles travel along ``rails" located in the center of the lane,
never deviating from the center as real human drivers do. 
In addition, CARLA's pedestrian models move mechanically according to mathematical gait functions.
Due to the complexity of human behavior and movement characteristics, pedestrians are particularly difficult to simulate.

These limitations point to the need for other solutions to obtain such data.
Virtual Reality (VR) offers a solution for collecting pedestrian data
% not only pedestrian trajectory
% but also body pose data 
in a safe environment free of real automotive dangers;
VR can be used to immerse a pedestrian in a virtual scenario analogous to one found in real-life,
and the pedestrian can be asked to cross a virtual road, or traverse the virtual environment, all the while their trajectory data is collected.
In addition, data collected from VR is collected from real humans, and thus does not suffer from the same human behavioral sim2real gap as simulated data. 
However, VR may exhibit limitations to realistic data collection.
For example, the VR user experience could be unrealistic, 
whether it's because of the graphics of the virtual environment, lag or delay in the VR view refresh frame rate,  or the weight of the VR headset on their head. 
% \aaditya{I know you don't want to necessarily talk at length about limitations, but to me the limitations that came to mind (so maybe replace one or more of the above with these) are the lack of nonvisual environmental cues? e.g. sounds, rush of air. All falls under the umbrella of ``experience could be unrealistic''. Also genuine question -- how's the peripheral vision on the VR headsets you use? I know from convos @Meta that this is often a big issue?} \marisa{to add on to limitations that come to mind...people who know they're observed act differently(ie perhaps more safely/predictable) than when unobserved}
These issues may interfere with the pedestrian's sense of immersion, and accordingly, 
may alter their actions and movements from what they would do on a real street.
Evaluating a VR system~\cite{anon2023} that was developed concurrently with this work through empirical studies,
% that is capable of collecting trajectory data in a wide variety of vehicle-pedestrian interaction scenarios, 
% including jaywalking, intersection-crossing, and walking alongside traffic on a heavily-trafficked road.
% such as jaywalking, unique environments, and interactions with agents in close proximity to one another.
we show that although users do not experience perfect sense of presence or immersion within the system, 
data collected from the system is mostly indistinguishable from that collected in real life,
providing evidence for our claim that the VR system is effective at collecting a genuine pedestrian response to vehicle motion. 
Our contributions are as follows:

\begin{enumerate}
    \item Through semi-structured interviews with 4 professionals in the autonomous vehicle (AV) field, 
    we show that a virtual-reality-based system for collecting pedestrian data in 
    traffic scenarios can address deficiencies in existing vehicle-pedestrian datasets 
    and limitations to existing real-life data-collection methods.
    \item Through an empirical quantitative and qualitative questionnaire of 62 individuals testing our system, 
    we demonstrate that such a virtual reality (VR) system can elicit a genuine human response to vehicle motion, 
    and result in pedestrian behaviors that mimic those of pedestrians interacting in real-life vehicle-pedestrian scenarios.
    \item Through an empirical survey of 290 individuals each answering 3 forced-choice questions comparing the realism of 2D trajectories collected in the VR system vs. in real-life, we measure the similarity of the VR data to real-life data.
    % \item We verify the usefulness and applicability of such a system using a statistical comparison of gait and walking patterns in the virtual environment and in real-life.
\end{enumerate}

% The remainder of this paper is organized as such: Section~\ref{sec:bkg} is Background and Relate

\section{Background and Related Work}
\label{sec:bkg}
% In this section, we establish background and describe related work.
\subsection{Human Behavior in Virtual Environments}
Many works have studied human behavior within virtual environments. 
Some works study the efficacy and quality of the VR experience via objective measures such as stride length and gait~\cite{Janeh2018-jr} as well as subjective measures such as subjects' ratings on presence questionnaires ~\cite{Drey2021-ek,Berkman2019-qp,Slater1997-pg,Stanney2003-ms}.
Some study specific behaviors of interest like normal walking~\cite{Nilsson2018-qi}, evacuating a building~\cite{Arias2022-ue}, collision-avoidance ~\cite{Berton2019-wu}, proxemics and group behavior~\cite{Podkosova2018-za,Norouzi2019-eb}, etc. Some focus on behaviors in specific professional areas like surgery~\cite{McCloy2001-pw,Li2017-td} or education~\cite{Bell_undated-sb}. Many methods evaluate the efficacy of VR systems by comparing human behavior in VR vs. real environments~\cite{Merino2020-fk}. Finally, other works evaluate how different conditions of the VR system's features affect user sense of presence behavior, such as level of photorealism of the environment rendering~\cite{Zibrek2019-hx}, locomotion methods~\cite{Tan2022-oz,Nilsson2015-uq,Olivier2018-ut}, and avatar appearance~\cite{Samaraweera2013-an,Ogawa2020-jf}. 
% \aaditya{last sentence a bit wordy but it's probs fine -- if you have time maybe reword/split up.} \marisa{I agree, there also isn't parallel sentence structure, so it's a little harder to parse.}

\subsection{Pedestrian Behavior Studies} 
Extensive study has been done to characterize pedestrian behavior in traffic environments and responses to vehicle behaviors~\cite{Jayaraman2018-he,Colley2020-ea,Lee2019-mr,Feng2021-oz,Weber2019-em,Chen2020-zr}.
Some works study AV behavior or evaluate improvements to the AV-pedestrian communication interface such as intent signalling methods~\cite{Jayaraman2018-he,Colley2020-ea,Lee2019-mr,Feng2021-oz,Lee2019-mr,Fuest2020-kl,Ackermans2020-xj,Deb2020-xx,De_Clercq2019-ot}.
Many of the aforementioned works, as well as others, use immersive VR environments to simulate scenarios of interest and study pedestrian's responses~\cite{Nunez_Velasco2019-ch,Schneider2020-wi,Bhagavathula2018-qn,Bhagavathula2018-qn, Feldstein2020-rd,Mahadevan2019-ns,Tran2021-rr,Nunez_Velasco2019-ch}.
Some works focus on evaluating the efficacy of VR systems for use in studying human behavior or evaluating safety systems or vehicle-pedestrian communication systems by comparing subject's responses in virtual environments to that of real-world environments~\cite{Nascimento2019-sx,Deb2017-dp,Schneider2020-wi,Higuera-Trujillo2017-tk}.
Work on VR-based dataset collection and benchmarking tools similar to that studied in this work has also been done, such as \cite{Dalipi2020-qp}, but their focus is on testing vehicle-pedestrian communication interfaces rather than on collecting realistic vehicle-pedestrian interaction behavior.
% and requires a motion-capture suit to obtain full-body pose and other complicated devices. 

\subsection{2D Pedestrian Trajectory Data Capture: Real-Life, Human-Informed, and Fully-Synthetic} 
Previous works have used various methods to capture vehicle-pedestrian interaction datasets.
The most popular datasets, naturally, are those collected in real-life environments
~\cite{Pellegrini2009-xp, Lerner2007-zo,Robicquet2016-ol}.
% As they are pedestrian-only, they lack vehicle-pedestrian interactions important to learning vehicle-pedestrian dynamics.
PedX~\cite{Kim2018-om} and PIE~\cite{Rasouli2019-yy} contain pedestrian trajectories collected at intersections, which is the closest data to our use case.
% annotated as bounding boxes in traffic environments, but no finer-grained pose information.
% pedestrian full body pose trajectories interacting with vehicles in traffic scenes; 
However, they too are limited in scope; PedX contains full body pose data but in limited environments such as crosswalks, 
and PIE contains only 2D bounding box annotations.
Popular open-source vehicle datasets 
also contain pedestrians~\cite{Chang2019-jr,Wilson2023-yc,Caesar2019}.
However, as verified in the first part of our study in Section~\ref{sec:pt1}, 
these datasets lack diverse pedestrian behavior and diverse vehicle-pedestrian interactions.
Most vehicle-pedestrian interactions in these datasets are at intersection crosswalks.

% Human pose datasets contain more fine-grained data on human skeleton motion in a variety of actions and activities. 
% Notable ones include Human3.6M~\cite{Ionescu2014-wu}, HumanEva~\cite{Sigal2010-ih}, 
% DensePose~\cite{Guler2018-kx}, and AMASS~\cite{Mahmood_undated-dj}; but there are many 
% others~\cite{Von_Marcard2018-oi,Varol2017-lt,Lassner2017-to,Andriluka_undated-le,Joo2019-cy}.
% None of these pose datasets contain extensive data in traffic settings, 
% and thus do not have many vehicle-pedestrian interactions\marisa{comma here?} if at all.

Some datasets are completely synthetic, collected autonomously via simulated environments that operate based on rules.
While some simulators only produce simplistic point trajectory behavior~\cite{Biswas2021-xy,Kothari2022-hl}, others are extensively customizable and controllable 
worlds that can model complex sensor data~\cite{Dosovitskiy2017-oy,Chen2021-yg,Yang2023-dk}.
Although simulators simplify the process of collecting large-scale data in diverse environments, there is a domain gap between synthetic and real data that is often difficult to measure, and many simulators do not reproduce human behavior well.
Because AV models ultimately must work for real pedestrians and vehicles, in this way synthetic data falls short.

Some methods, such as 
driving simulators~\cite{Silvera2022-uk,Michalik2021-bs}, invite human input to elicit real human behavior.
The Garden of Forking Paths~\cite{Liang2019-xo} collects annotations of trajectory continuations from human annotators
to create a multi-future pedestrian trajectory dataset in simulated environments reconstructed from real-world scenes.
However, they only annotate trajectories at the 2D point-level, through a very limited keyboard and mouse interface, a far cry from the controllability of a VR headset and environment.

\section{Part 1: Understanding Limitations of Existing Vehicle-Pedestrian Datasets}
\label{sec:pt1}
In order to better substantiate the use cases for a VR pedestrian data-collection system, we sought to confirm the apparent limitations of existing vehicle-pedestrian interaction datasets with AV professionals.
Thus, in the first part of our study, we conducted semi-structured interviews with professionals 
who had experience working with vehicle-pedestrian trajectory datasets.
Our goal was to answer the research question:
\textit{What are the limitations of existing vehicle-pedestrian datasets with respect to the availability and quality of pedestrian data?
}

\subsection{Semi-Structured Interview Recruitment and Design}
For the interviews, we recruited 3 academic researchers and 1 industry practitioner through direct contacts. Two academic participants have performed research in the AV perception, prediction, and planning stack (referred to by I1 and I2).
The other academic participant performs research in social navigation 
for mobile robots that interact closely with humans (I3).
The industry participant, meanwhile, is a systems test engineer at a large AV company (I4).
During the study, all participants self-reported that they had experience working with vehicle-pedestrian interaction data.

%Before the interview, we asked participants to not reveal any confidential or identifiable information about themselves or their affiliations, as well as that we would anonymize responses and transcriptions. We also told them any quotes included in this work would be vetted for their approval before being published. The study was also approved by our institution’s IRB procedures.

In our interviews, we first asked participants to describe their area of expertise;
Then, we asked them to describe their understanding of the current state of vehicle and pedestrian trajectory datasets:
\textit{Which trajectory datasets you have worked with?
What do you feel are the current limitations of these datasets?}
We probed deeper into the state of current vehicle-pedestrian trajectory datasets with questions such as 
\textit{What kinds of scenarios are lacking in real datasets?} and 
\textit{What existing methods are there for improving performance in uncommon or out-of-distribution scenarios?}
Finally, we asked directly for their opinion about the potential limitations and benefits of data collected in a virtual environment via a VR simulator.
We asked questions such as \textit{Given a Virtual Reality system
in which a pedestrian walks around while wearing a VR headset in a virtual traffic environment while sensors capture their movements:
what are the potential benefits and limitations of such a system?}

% KK: Seem unnecessary
%After completion of the interview, participants were compensated \$20 for their participation, and their quotes were used with permission in this manuscript.

\subsection{Interview Responses}
To analyze the state of the limitations of current trajectory datasets, we used inductive thematic analysis~\cite{Braun2006-vi}, grouping the themes that arose during the interviews into four distinct categories, which we describe below.

\subsubsection{Lack of Uncommon but Important Scenarios}
Vehicle-pedestrian datasets lack scenarios that are important to AV safety,
but rare in-the-wild or difficult to collect data for. 
% Datasets may contain plenty of pedestrians at crosswalk intersections, but data for jaywalkers is limited. 
% Understanding jaywalker behavior is essential to pedestrian safety, even though they are sparse in real-life.
% Unfortunately, it is dangerous and unethical to direct pedestrians to jaywalk in real-life or even in constructed real environments.
I2 commented that vehicle-pedestrian datasets contain plenty of data of pedestrians crossing roads at intersections, 
but little data of pedestrians walking on the sidewalk alongside the road.
These scenarios are just as important as pedestrians crossing at crosswalks, 
because pedestrians walking alongside the road could become jaywalkers in the near future.

% VR can be used to simulate these edge-case scenarios that are too difficult or too dangerous to recreate in real-life.

\subsubsection{Lack of Interesting Vehicle-Pedestrian Interactions}
Both I1 and I2 pointed out that the most popular public vehicle-pedestrian datasets lack ``interesting interactions." 
One of the most popular datasets~\cite{Chang2019-jr, Wilson2023-yc} in particular contains a subset that was
compiled specifically for the ``interestingness" of its 
scenarios based on internally-defined heuristics,
such as number of other agents present in the scene, speed changes, and lane changes. 
However, I1 remarked that these scenarios were still mostly uninteresting,
claiming that it is difficult to come up with a straightforward 
heuristic to separate "interesting" scenes from "boring" scenes. 
I2 stated that around 75-80\% of the data they worked with is ``uneventful",
and that there was insufficient diversity in scenes.
Scenes often featured empty or minimally populated roads where only the ego-vehicle, the vehicle performing the data collection, was present.
Vehicles also lacked varied behaviors such as lane changes. 
I3 claimed that the pedestrian-only datasets they worked with also lacked variety in environment layouts 
and did not contain many pedestrian interactions.

\subsubsection{Lack of Fine-grained Trajectory Features}
Vehicle-pedestrian datasets lack richness in representation.
I1 commented that existing trajectory datasets are heavily preprocessed from raw lidar data, 
losing a lot of the richness of the raw representation.
Popular trajectory datasets such as Argoverse~\cite{Chang2019-jr,Wilson2023-yc} and NuScenes~\cite{Caesar2019-az} represent pedestrian trajectories as 2D points rather than full 3D human bodies. 
I2 notes that while pedestrian body orientation is also recorded, it still falls far from the detail available in full body pose. 
In tracking datasets, agents are often annotated with simple 3D rectangular bounding boxes. 
I3 explained that pedestrian bounding boxes often overlap with one another, because humans occupy their bounding boxes only sparsely.
Overlapping bounding boxes appear to be colliding; this may lead to model training issues such as failure to understand collision-avoidance.
Thus, bounding boxes also fall short in representing the richness of pedestrian motion.

Pedestrian body skeleton pose and head direction can inform understanding of pedestrian intent high-level pedestrian motion. 
For example, pedestrians often look both ways before they cross the street, and a pedestrian that is looking in the opposite direction of an oncoming car is less likely to stop for that car, as they do not see it. 
In addition, traffic officers often stand in the middle of the road to direct traffic with their arm movements.
However, as I4 pointed out, their 
company's AV system has limitations in predicting the behavior of traffic officers, often mis-predicting them as pedestrians crossing the road
rather than static pedestrians who will not intercept the vehicle's path in the near-future.

% Thus, for both pedestrians and vehicles, knowing cues such as turn signals and 
% driver gaze and hand motions may lead to better vehicle intent prediction.

% Human skeleton body pose is a representation more compact than raw lidar or camera data, 
% but more rich than simple 2D point trajectories or bounding boxes.
% It is difficult to extract accurate skeleton poses from monocular camera data or lidar data in public datasets.
% However, in the VR system we evaluate in this paper, 
% multiple cameras set up at different angles are used to robustly 
% and accurately extract body skeleton from camera data,
% enabling the capture of body pose and eye gaze data 
% that is much more difficult to obtain with online AV setups.
% Marisa: this is kinda cool info, kinda hope it shows up somewhere else

\subsubsection{Summary}
The shortcomings of current datasets 
point to the need for additional data solutions to supplement existing datasets. 
The VR pedestrian simulator we evaluate in this paper can be manipulated to recreate diverse and uncommonly seen traffic scenarios, addressing the first two themes.
It can also record body pose information,
addressing the third theme.
There were some concerns brought up by our interview participants 
that data collected from a VR system may have a domain gap with real-world data, just as other simulators do.
In the next two sections, we provide evidence that in spite of the 
domain gap,
the VR environment still elicits a genuine and realistic pedestrian response. 

\section{VR Pedestrian Trajectory and Pose Collection System Overview}
We briefly describe the VR pedestrian data-collection system that we evaluated.

The simulator uses the HTC VIVE Pro 2%~\cite{noauthor_undated-ga} 
as its VR system. Steam VR%~\cite{noauthor_undated-rg}
is used to interface the HTC VIVE devices with a controllable game engine, 
so they can be used as input devices within the simulation.
The simulator is built on top of the CARLA simulator%~\cite{Dosovitskiy2017-oy} 
using Unreal Engine.%~\cite{noauthor_undated-ci}. 
Since Steam is only available on Windows machines and the HTC 
VIVE Wireless Adapter transmitter must be connected to the computer via an RP-SMA port,
the Alienware Aurora R13 Gaming Desktop%~\cite{noauthor_undated-de} 
is a suitable machine for use as the underlying server.
The system specifications are as follows: 12th Intel core i9 CPU, NVIDIA RTX 3080 GPU, 64 GB memory, M.2 SSD 2TB storage, Windows 11 Home OS.

\begin{figure}[t]
  \includegraphics[width=\columnwidth]{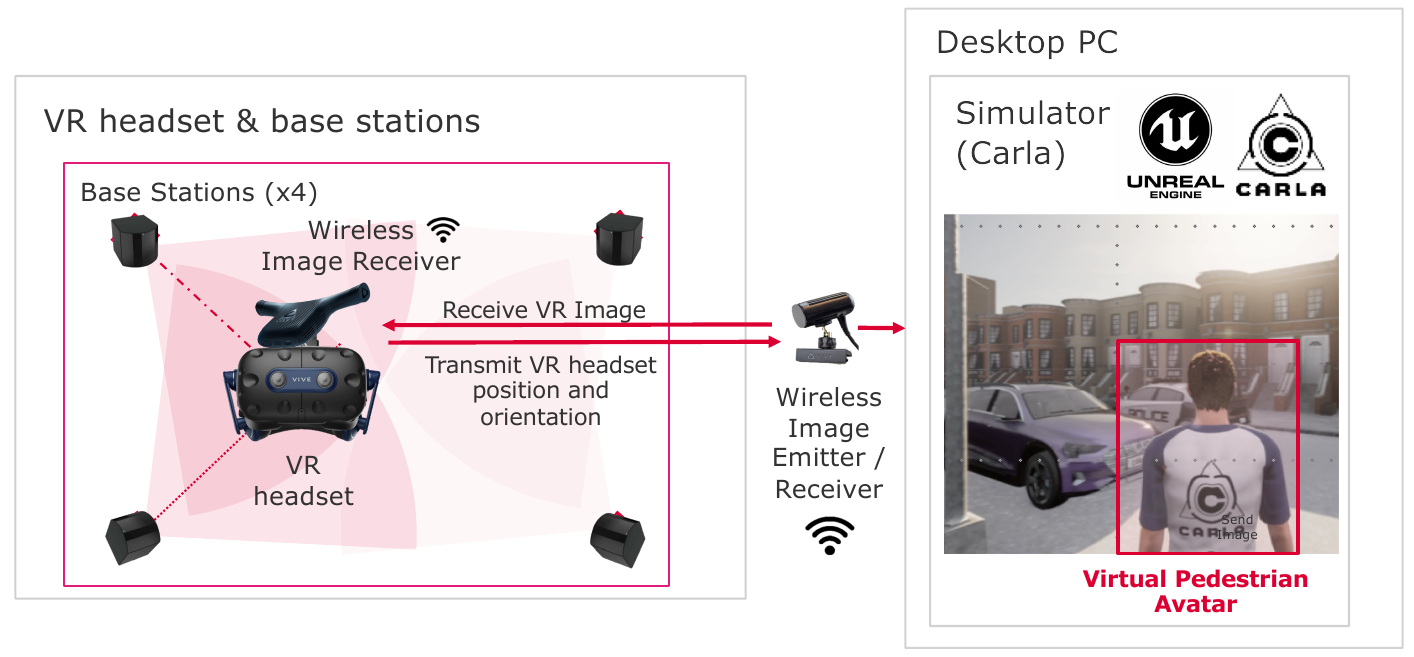}
  \caption{Schematic of the VR data collection system.}
  \label{fig:system_schematic}
\end{figure}

\begin{figure}[t]
  \includegraphics[width=\columnwidth]{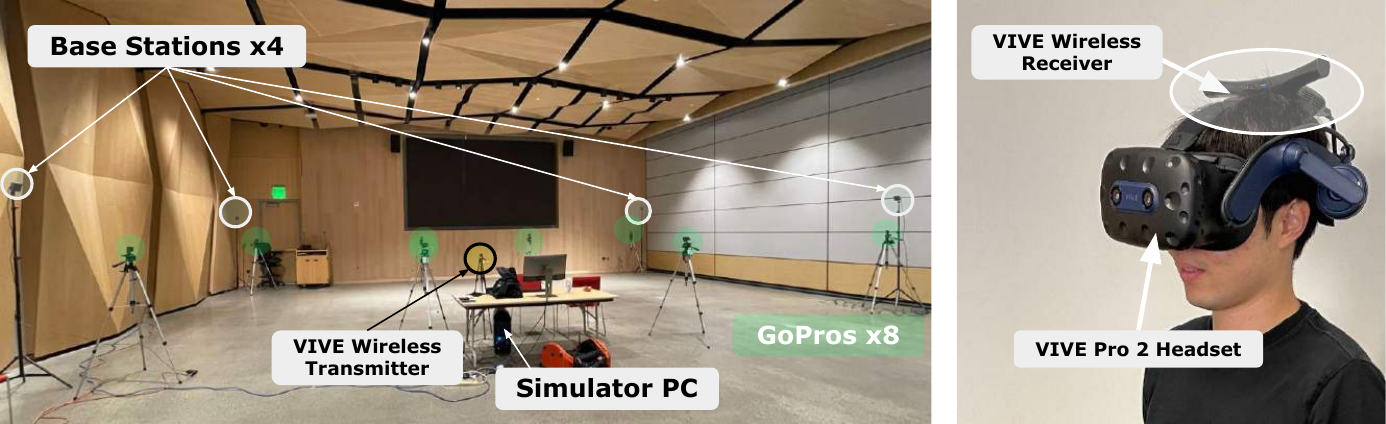}
  \caption{Left: The VR system set up in the classroom data collection space.
  % The PC running the simulator underneath the table in the center of the image, and the VIVE VR wireless signal receiver mounted on a small tripod behind the table.
  Right: A pedestrian subject wearing the VR headset.}
  \label{fig:system}
\end{figure}

The data collection environment requires a rectangular collection space free of obstacles measuring approximately 40'x 20' in dimensions.
Four HTC VIVE base stations are set up in the four corners of this space.
Eight GoPro Hero 10%~\cite{noauthor_undated-mh}
cameras are set up along the two long edges of the space, which record the subject from different angles so that their skeleton pose can be extracted out later.
A schematic of the VR system components is shown in Figure~\ref{fig:system_schematic}, and the actual data collection space with system setup is shown in Figure~\ref{fig:system}.

Because the engineering and design of the VR pedestrian simulator is not the primary contribution of this paper but of~\cite{anon2023}, we leave further details about the data collection system to that work.

\section{Part 2: VR Sense of Presence User Study}
To determine the degree to which users felt immersed and present in the virtual world, 
% the VR pedestrian simulator as a new data source for vehicle-pedestrian interactions to supplement existing datasets, 

\subsection{Study Design}
We recruited participants by word-of-mouth, university email lists, 
and physical flyers posted around the university campus.
Participants with physical disabilities, vulnerable subjects, and minors were excluded from participation.
Participants came to the university campus to participate in the study, 
where we set up the data collection environment as described in the previous section in a large, empty classroom. 

\begin{figure}[t]
  \includegraphics[width=\columnwidth]{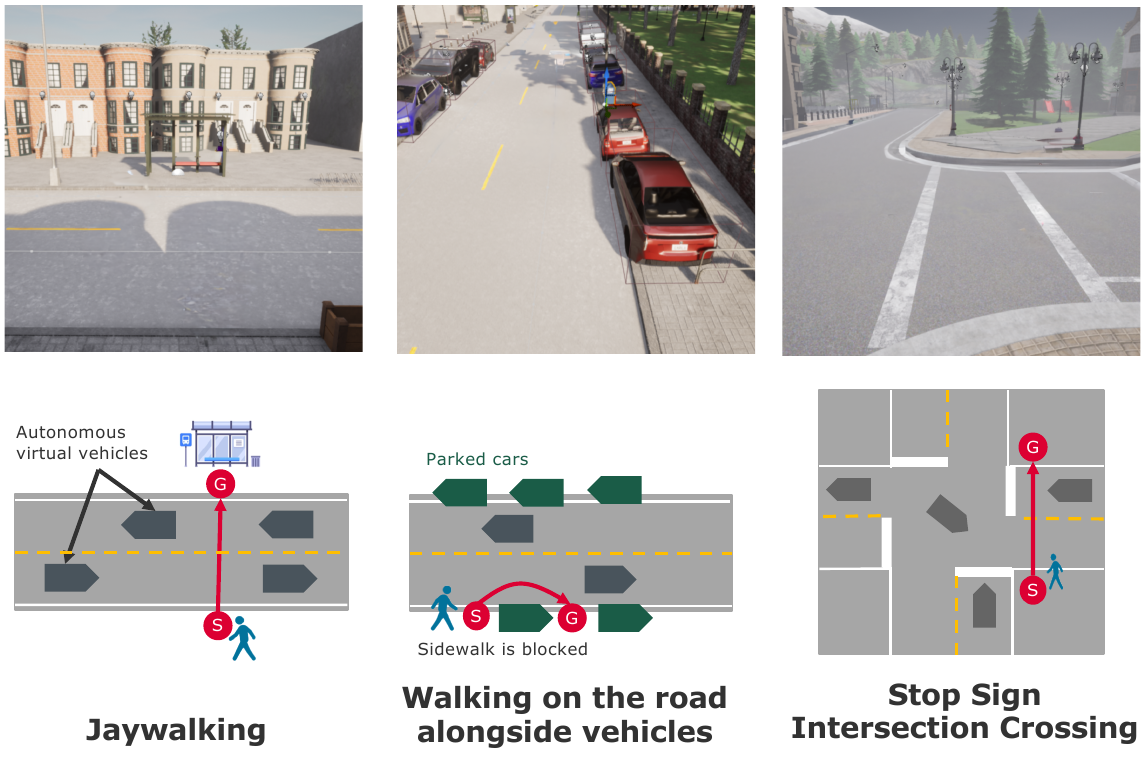}
  \caption{The three tasks that we ask our study participants to complete. Top: snapshot of each task as seen in the CARLA virtual environment. Bottom: schematic of each task; ``S" denotes participant start location and ``G" denotes goal. Schematic images also used in \citet{anon2023}.}
  % \Description{}
  \label{fig:scenarios}
\end{figure}

After signing an informed consent form, participants were asked to put on the VR headset
and to familiarize themselves with the VR headset and virtual environment by walking 
around in a version of the simulation without moving vehicles.
Then, we asked participants to complete 3 tasks, 
each of which featured a different traffic environment 
in which they had to walk to reach a goal destination while in close interaction with moving vehicles: 1) jaywalking across a two-way street, 
2) walking alongside moving vehicles on the road, 
and 3) crossing a crosswalk at a 4-way intersection with stop signs.
The 3 tasks are depicted in Figure~\ref{fig:scenarios}.
To inform participants of their goal destinations, 
we placed colored square markers on the ground within the virtual environment, and used commands such as
``Do you see the colored square on the ground on the other side of the road? Please walk to it" to direct the user.
After completing all 3 tasks, participants were asked to complete a presence questionnaire to evaluate their experience.

% Participants received \$20 compensation for approximately 20 minutes of their time
% to experience the VR system, have their data collected,
% as well as take the post-condition questionnaire.
% This study was approved by the university's IRB procedures.

\subsection{Self-Reported Measures}
To design the questions asked in the post-experience questionnaire, 
we used a combination of questions derived from previous works, 
questions modified from previous works, 
and custom questions designed specifically to evoke defining attributes of the VR user experience.
We designed 12 quantitative questions using a semantic differential scale~\cite{Tina_Ovad_undated-dj} from 1 to 7,
as well as 4 free-response questions with free text entry.
We grouped the questions into 3 categories reflecting different aspects of trustworthiness of the data collection system: 
sense of presence experienced within the virtual environment~\cite{Berkman2019-qp,Witmer1998-tt} (coded by the letter P), 
sense of agency (A), 
and behavioral and experiential similarity to real-life (B).
In total, the post-condition questionnaire was composed of 16 questions
aimed to cover a range of subjective ratings
while keeping the time for participants to complete the questionnaire within 10 minutes.
Questions, sources, categorizations, and semantic differential anchors for the questions are recorded in Table~\ref{tab:quest}.

Participants also reported various demographics: age, gender, 
VR experience (5-point Likert-type scale), video game frequency (5-point Likert-type scale),
frequency of jaywalking behavior (5-point Likert-type scale),
and level of alertness (Stanford Sleepiness Scale~\cite{Shahid2012-nm}).
Summary of participant demographics is reported in Table~\ref{tab:demo}.

\begin{table*}[ht]
    \centering
    \scriptsize
    \begin{tabular}{p{.1cm}p{.5cm}p{.7cm}p{9.8cm}p{4.8cm}}
    \hline
    % \smallskip
        \textbf{ID} & \textbf{Source} & \textbf{Category} & \textbf{Question} & \textbf{Semantic Differential Anchors (1$\rightarrow$ 7)} \\
    \hline
   P1 & SUS~\cite{Usoh2000-de} & presence & Please rate your sense of “being there” in the environment. & did not have a sense of "being there" $\rightarrow$ normal experience of “being there” \\
   P2 & SUS~\cite{Usoh2000-de} & presence & When you think back on your experience, do you think of the environment more as images that you saw, or more as somewhere that you visited? & Images that I saw $\rightarrow$ somewhere that I visited \\
   P3 & WS~\cite{Witmer1998-tt} & presence & How real did the objects in the environment seem? & very fake, clearly images $\rightarrow$ very real, like I could touch them \\
   P4 & \textit{custom} & presence & During the VR experience, were you more concerned with the real world (this classroom) or the virtual world? & real world $\rightarrow$ virtual world \\
   P5 & SUS~\cite{Usoh2000-de} & presence & When you think back on your experience, do you think of the vehicles more as images that you saw, or as things you interacted with? & Images that I saw $\rightarrow$ things that I interacted with \\
   P6 & \textit{custom} & presence & During the experience, did you often think to yourself that the vehicles were physical objects that could have actually hit you and caused you injury? & very much so $\rightarrow$ not at all \\
   A1 & \textit{custom} & agency & How comfortable did you feel moving around in the environment? & very uncomfortable $\rightarrow$ very comfortable \\
   A2 & AE~\cite{Gonzalez-Franco2018-kl} & agency & How much did you feel like you could control the virtual body? & did not feel much agency $\rightarrow$ could control it like own body \\
   A3 & SPES~\cite{Hartmann2016-pv} & agency & How freely did you feel you could move in the environment? & movements were restricted $\rightarrow$ movements were free \\
   B3 & NASA-TLX~\cite{Hart1988-yz} & behavior & How mentally demanding were the tasks compared to doing them in the real world (e.g. on a real street)? & less demanding $\rightarrow$ more demanding \\
   B1 & \textit{custom} & behavior & Did you feel your head, arm, and body movements were the same as they would have been in the real world? & completely different $\rightarrow$ exactly identical \\
   BF1 & \textit{custom} & behavior & What parts of your movements were different than how they would have been in the real world? & \textit{free response} \\
   B2 & \textit{custom} & behavior & Did you feel your decisions about when to act were the same as they would have been in the real world? & completely different $\rightarrow$ exactly identical \\
   BF2 & \textit{custom} & behavior & What parts of your decisions were different than how they would have been in the real world? & \textit{free response} \\
   PF1 & \textit{custom} & presence & What aspects of the systems or environment were realistic? & \textit{free response} \\
   PF2 & \textit{custom} & presence & What aspects of the system or environment were unrealistic? & \textit{free response} \\
    \hline
    \end{tabular}
    \caption{Post-condition questionnaire used in the study.}
    \label{tab:quest}
\end{table*}

\begin{table}[t]
    \centering
    \footnotesize
    \begin{tabular}{ll}
    \hline
    \smallskip
        \textbf{Variable} & N=64 \\
    \hline
    Age \{18-37\} (years) & 24.71 (4.17) \\
    \% Male & 63\% (n=39)\\
    \% Female & 37\% (n=23)\\
    Video gaming frequency \{1-7\} & 3 \{1-5\} \\
    VR experience \{1-7\} & 2 \{1-4\}\\
    Jaywalking frequency \{1-7\} & 3 \{1-5\} \\
    Level of alertness \{0-7\} \cite{Shahid2012-nm} & 6 \{3-7\}\\
    \hline
    \end{tabular}
    \caption{Summary of participant demographics. Continuous variables are summarized as \textit{mean} (\textit{standard deviation}) and ordinal variables are summarized as \textit{median} \{\textit{range}\}.} \label{tab:demo}
\end{table}

\subsection{Results and Analyses}
\begin{figure}[t]
  \includegraphics[width=\columnwidth]{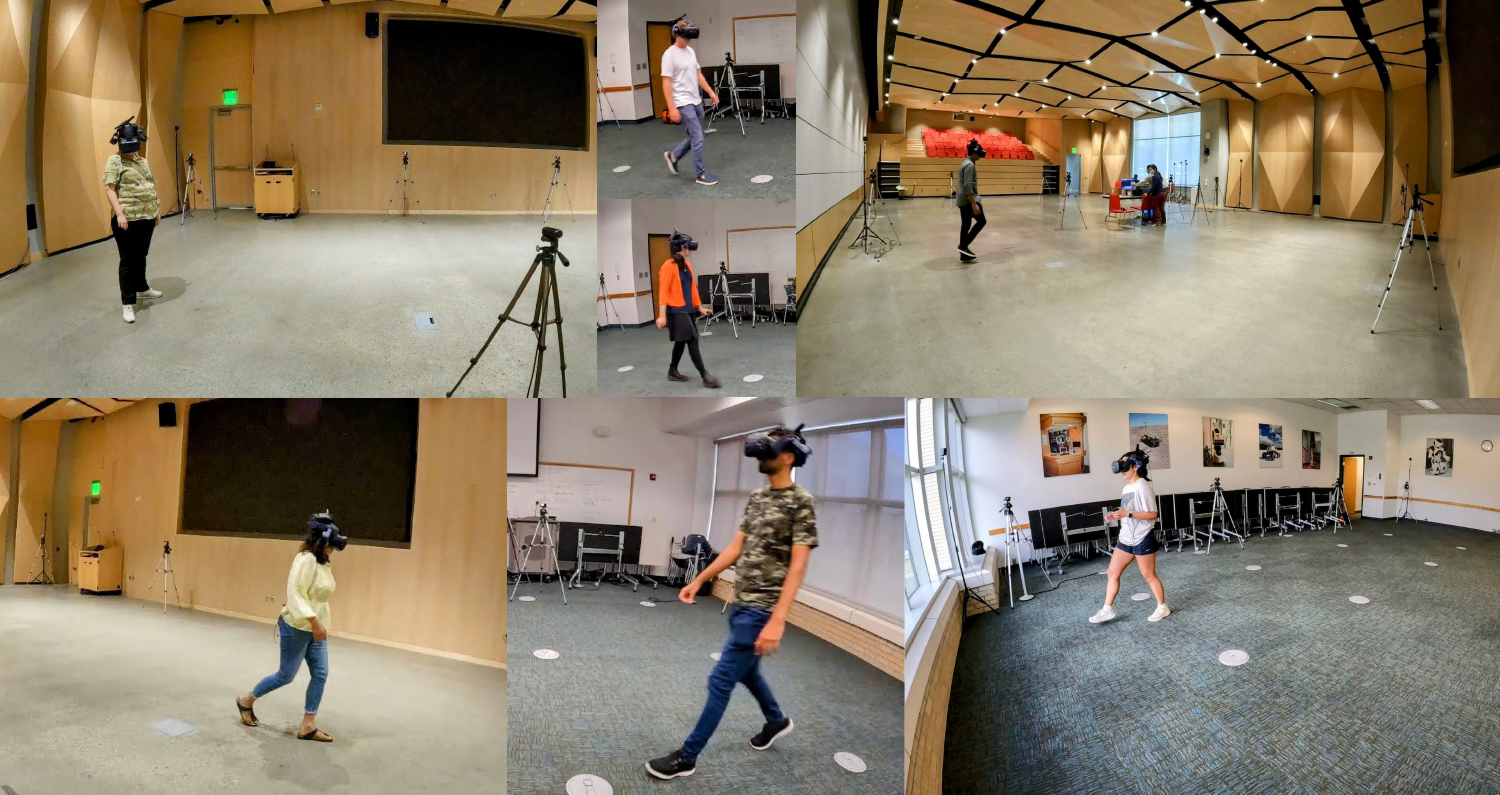}
  \caption{Various study participants wearing the VR headset and walking around in two different classroom-setting data collection environments.}
  \label{fig:people}
\end{figure}

\subsubsection{Quantitative Measures}
A total of \texttt{N=62} participants participated in our study. Aggregating all responses across all questions in a category, users rated their sense of presence (P) 5.6, sense of agency (A) 5.5, and similarity of their behavior and experience to that of the real world (B) 5.3. These numbers suggest that users experience a relatively moderate-high sense of presence, agency, and behavior similarity to the real world. 
The users' aggregate response of 5.3 in the behavioral similarity category is a little lower than that in the sense of presence category and sense of agency category, possibly suggesting that high sense of presence is not sufficient to guarantee that a pedestrian behaves exactly the same in VR as they do in the real world. One possible explanation for this is that, for the high proportion of users who rarely used VR, the novelty factor of the VR environment causes them to behave differently, even though they are experiencing a high degree of presence and immersion.

% Another possible explanation for the small disparity in scores between P/A and B is that complete realism of the virtual environment may not be necessary to experience evoked by the qualitative results below,

We report the average user ratings for each question in Table~\ref{tab:user_study}.

% The behavioral similarity questions (B) provide a quantitative comparison of the user's subjective experience in the VR system with that in reality. 
% A score of 7 would mean participants feel that their behavior and experience in the VR system is completely similar to that in the real world. 
% A score of 1 would mean that they feel they are completely different.
\begin{table}[t]
    \centering
    \resizebox{\columnwidth}{!}{%
    \begin{tabular}{p{1.2cm}|cccccc|ccc|ccc}
        \hline
        \noalign{\smallskip}
                 \textbf{Question Category} & \multicolumn{6}{|c|}{\textbf{Presence}} & \multicolumn{3}{c|}{\textbf{Agency}} & \multicolumn{3}{c}{\textbf{Behav. Sim.}} \\
        \noalign{\smallskip}
        \hline
        \noalign{\smallskip}
        \textbf{ID (as in Table~\ref{tab:quest})} & \textbf{P1} & \textbf{P2} & \textbf{P3} & \textbf{P4} & \textbf{P5} & \textbf{P6} & \textbf{A1} & \textbf{A2} & \textbf{A3} & \textbf{B1} & \textbf{B2} & \textbf{B3} \\
        \noalign{\smallskip}
        \hline
        \noalign{\smallskip}
        \textbf{Rating} & 5.9 & 5.3 & 5.0 & 6.2 & 5.9 & 5.3 & 5.1 & 5.6 & 5.9 & 5.0 & 5.3 & 4.8 \\
        \noalign{\smallskip}
        \hline
        \noalign{\smallskip}
        \textbf{Average} & \multicolumn{6}{|c|}{\textbf{5.6}} & \multicolumn{3}{c|}{\textbf{5.5}} & \multicolumn{3}{c}{\textbf{5.3}} \\
        \noalign{\smallskip}
        \hline
    \end{tabular}
    }
    \caption{VR Presense Questionnaire Evaluation Results.}
    \label{tab:user_study}
\end{table}

\subsubsection{Qualitative Measures}
The qualitative responses evoke greater detail from the participants' experiences.
In response to the free-response question, \textit{What aspects of the systems or environment were realistic?}, participant responses fell into the following major categories:
\begin{enumerate}
  \item Traffic patterns and flow were very authentic.
  \item The visual rendering contributed to a realistic experience: the relative sizes and dimensions of cars, buildings, roads, sidewalks, trees and other aspects of the environment felt accurate, appropriately sized, and well-rendered. The environment felt realistically designed. Some users commented that the curb appeared so realistic that it caused them to stumble when they actually tried to step onto it, as no physical curb existed in real-life.
  \item Movement of images along the line of vision was smooth when the user moved their head or body.
  \item Cars moved and behaved realistically with respect to speed, positioning, and timing.
  \item The environment elicited emotions and caution similar to real-world experiences, such as feeling threatened by cars and being conscious of making mistakes, like crossing the road at the wrong time.
\end{enumerate}

In response to the question, \textit{What parts of the system or environment were unrealistic?}, 
the main concerns brought up by participants include:
\label{sec:pt2_realism}
\begin{enumerate}
    \item Dizziness and nausea while using the system
    \item System stability issues: lag, flicker, and instances in which the virtual space would re-calibrate and the participant would be transported within the virtual environment to a different location despite not having moved
    % However, this is not so much of an issue as the visual perception did not change
    \item Awareness of the real world and obstacles in the real world, which took away from sense of presence
    \item Lack of peripheral vision due to the construction of the VR headset goggles, which resulted in participants turning more to the left and right to check for vehicles than they do in real life
    \item Lack of environmental sound
    \item Imperfect visual information: lack of stop signs and walk signals at intersections, game-asset quality of buildings and surroundings, not all body parts visible in the simulation
    \item Imperfect tactile information: sidewalk curb observed in virtual environment, but real-world environment lacked a tactile height change
    \item Mechanical and unnatural movement of vehicles, which did not deviate from fixed paths nor yielded to pedestrians
    \item Lack of drivers in vehicles, which rendered users unable to make eye contact with drivers to determine when to cross
    \item Lack of other pedestrians in the environment
    \item Being told where and when to start and stop, unlike in real life where this is self-determined
\end{enumerate}

In response to the question, \textit{What parts of your movements were different than how they would have been in the real world?} participant responses group into the following major concerns:
\begin{enumerate}
    \item Did not turn head as much due to weight of the VR headset and limited peripheral vision
    \item step movements differed due to difference between visual perception and tactile perception (such as the curb)
    \item Walking style: keeping hands in front of body as a defensive body posture to avoid bumping into walls in the real world, or to fend off aggressive vehicles in the virtual world
    \item Navigation around static objects in the virtual world such as parked cars differed due to the unexpected dynamics of those objects, such as moving aside when users collided with them
\end{enumerate}

In response to the question, \textit{What parts of your decisions were different than how they would have been in the real world?} subjects reported the following major themes:
\begin{enumerate}
    \item Some subjects were more hesitant to move in the virtual environment than in real life, due to the realism limitations discussed in the responses to the question above. 
    % factors such as that the vehicle movements differed from real life, vehicles were driver-less and thus did not signal their intentions to the pedestrians, the VR headset had restricted peripheral vision, VR system fidelity such as frame dropping and latency, and the weight of the VR headset slightly impeded natural head and body movement.
    \item Other subjects had an added sense of urgency to act, because of unexpected or erratic vehicle behavior, or because lack of peripheral vision limited the subjects' awareness of cars
    \item Some subjects exhibited more risky behavior due to less fear about being hit by a virtual car.
\end{enumerate}

The qualitative responses from our user study affirm that, while the system does have areas for improvement, there are noteworthy strengths. Participants largely found the visual aspects and their own emotional responses to be authentic to their real-world experience.
Some users did report aspects of the virtual environment that seemed unrealistic; yet sense of presence, agency, and behavioral similarity to real-life still seems high (Table~\ref{tab:user_study}). This seems to suggest that complete realism of the virtual environment may not be necessary to have high sense of presence,

Furthermore, for users that found their behavior or movements differing from that in real life, the different behaviors are not necessarily less genuine than those of real life.
Uncommon, edge-case scenarios sometimes do not feel real because they are unexpected, but they are no less important for ensuring the safety of AVs. 
The VR system is exactly designed to evoke pedestrian response in those edge-case scenarios, thus improving the collection of data in those scenarios.

\section{Part 3: Third-Person Realism Evaluation of VR vs. Real-Life vs. Synthetic 2D Trajectory Data}
First-person subjective evaluation provides valuable insights into how well the VR system evokes realistic behavior. 
However, to further substantiate the comparability of real-life and VR data, we perform an additional third-person evaluation evaluative study. 
In this study, we asked external evaluators to try to distinguish between trajectory data collected in real-life vs. VR. This study evokes a different facet of trajectory realism, one that comes not from first-hand experience, but from third-person evaluation.

\subsection{Study Design}
To collect real-life jaywalking trajectories, we visited a two-way single-lane street near the university that is well-trafficked by jaywalkers as well as vehicles.
We chose this setting for its potential to provide real-life trajectories in a similar geographic layout to that we used in the VR jaywalking scenario.
In contrast to the VR scenario, there were fewer vehicles as well as more irregular gaps between vehicles, making it easier for jaywalkers to find a gap in which to cross.
% It's important to note that while are a level of potential ``danger," 
% However, it's important to note that the scenarios at this real-life location did not feature as dangerous traffic scenarios as those in the VR jaywalking scenario; 
A DJI Mavic Mini drone~\cite{noauthor_undated-ln} was used for data collection, flown at a height of approximately 25 meters.
In a 1-hour time interval, we secured footage for 7 jaywalkers.

The trajectories of these jaywalkers were manually annotated via a trajectory annotation tool
and interpolated to 20 frames per second to match the frame rate of the trajectory data collected from the VR system. 
The resulting trajectories were smoothed via Gaussian smoothing with a standard deviation of $0.35$ seconds,
as the VR system also performs some smoothing to eliminate noise from the collected trajectories.

\begin{wrapfigure}{l}{0.5\columnwidth}
  \centering
  \includegraphics[width=0.5\columnwidth]{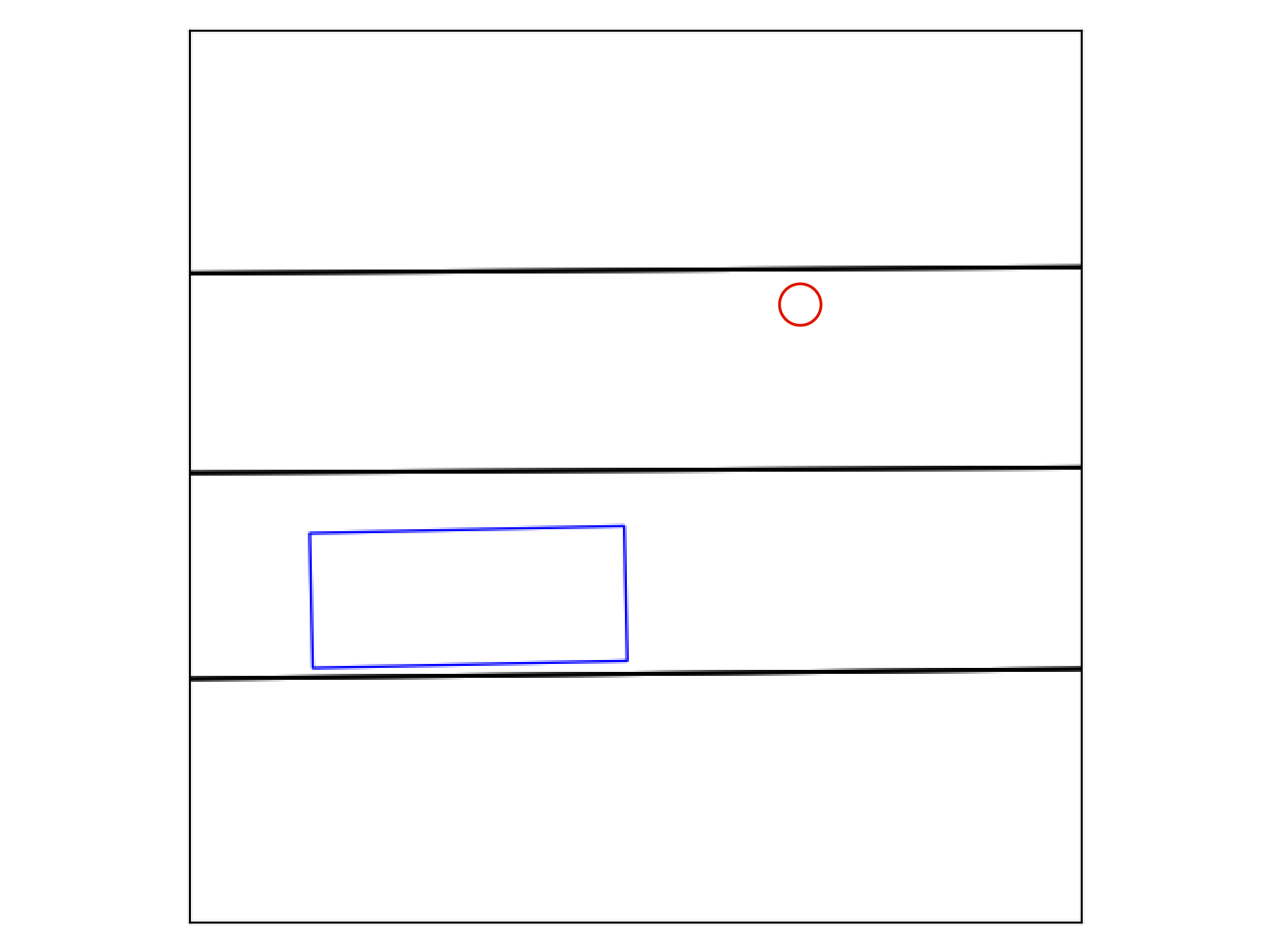}
  \caption{A frame from an example schematic shown to survey respondents.}
  \label{fig:2d_schem}
\end{wrapfigure}

\
The trajectories were rendered as animated gif images, in which the pedestrian was depicted as a small red circle, and vehicles as blue rectangles. 
A random rotation was added to each image to eliminate visual bias due to the layout of the scene. 
The spatial limits of the animation were set such that the users had an ample field of view to see oncoming vehicles from both sides (which is important for judging pedestrian behavior). 
All animations were set to have a similar field of view. 
For the real-life trajectories, only the single jaywalking pedestrian was annotated.
Scenes were chosen such that the pedestrians kept an ample distance from other pedestrians such that they did not visibly interact or influence each other's motion;
textit{i.e.}, the vehicle-pedestrian interaction would be the sole interaction evaluated.
An example schematic in the style of the animations is shown in Figure~\ref{fig:2d_schem}.

The survey consisted of 3 forced-choice questions featuring a comparison between either a real-life trajectory and a VR-collected trajectory,
a real-life trajectory and a synthetic piecewise-linear trajectory,
or a VR-collected trajectory and a synthetic piecewise-linear trajectory. 
Next to each pair, we prompted respondets with the question, \textit{Which one looks more real?} 
The synthetic piecewise-linear trajectory was included as a control; 
expectation is that users would  more easily perceive that trajectory as ``fake."
Pairings was selected at random, choosing one animation from 7 real-life trajectories, one from 8 VR trajectories, and one from 2 synthetic trajectories. 
With 10\% probability, the VR trajectory was swapped out for the synthetic trajectory, 
and with 10\% probability, the real-life trajectory was swapped out for the synthetic trajectory; 
thus, $\sim 20\%$ of pairings featured a synthetic trajectory in the comparison.
The other $\sim 80\%$ of pairings featured a comparison between real-life and VR trajectory. 
Display order was randomized.
Finally, there was an optional free-response question at the end: 
\textit{What characteristics did you use to distinguish between real and fake?}
The full set of animations for each of the three categories will be provided in the supplementary material.

The survey was distributed via word-of-mouth as well as by university mailing list. 
Response time for the survey was around 3 minutes per person.
Due to the low time-commitment nature of the survey, no compensation was offered for completion of the survey.

\subsection{Results and Analyses}
A total of 302 respondents answered the survey for a total of $ 302 \times 3 = 906$ forced-choice comparisons made by survey respondents.

Consistent with expectations about the ``control" variable, evaluators could easily tell the difference between a simple synthetic policy and a genuine human trajectory. 89.3\% of real-life / synthetic comparisons were evaluated in favor of the real-life trajectory looking ``more real" than the synthetic.
The response was similarly high for the VR / synthetic comparisons, with
88.9\% evaluated in favor of the virtual over the synthetic (Table~\ref{tab:3rd_person}).

Also consistent with expectations, the real-life trajectories were the most-frequently chosen as ``more real," with approximately 67\% of responses choosing it as more ``real" in a pair (Table~\ref{tab:3rd_person}).
However, VR trajectories do not lag far behind;
59\% of responses chose the VR trajectories as ``more real" in a pair (Table~\ref{tab:3rd_person}). 
Of particular note is that 36.5\% of real-life / VR comparisons were evaluated in favor of the VR trajectory looking ``more real" (Table~\ref{tab:pairings});
these responses claimed that the VR trajectory looked ``more real" than a real trajectory. 

Though 36.5\% is still a bit away from 50\%, which would mean that real-life trajectories are indistinguishable from VR trajectories, this number still substantiates the claim that VR trajectories can evoke genuine pedestrian responses.
First of all, the results confirm that real-life trajectories are less distinguishable from VR trajectories than they are from fully-synthetic trajectories, supporting the claim that the VR system evokes more natural pedestrian responses than simple policies used by synthetic pedestrian simulators~\cite{Biswas2021-xy,Kothari2022-hl}. 
Second, there could be other reasons evaluators are able to tell the difference between real and VR trajectories that have nothing to do with how ''real" VR trajectories are.
For example, one respondent who evaluated 3 comparisons between real-life and VR trajectories and marked all the real-life ones as more ``real" gave the free response answer that ``the fake pedestrians seem to cross dangerously close to the vehicles, and sometimes stop in the middle of the road..." 
Though this respondent labelled all real-life trajectories as ``more real" than the corresponding VR ones, their explanation for their rating is not necessarily true. 
It is generally considered dangerous and not commonly observed that pedestrians ``cross dangerously close to the vehicles." As observed in all drone-collected videos, the average jaywalker in real-life is a ``safe" jaywalker who only jaywalks when there is sufficient gap in the vehicles.
However, sometimes there are pedestrians who do engage in risk-taking behavior.
Sometimes there are jaywalkers who jaywalk even on heavily-trafficked roads. These jaywalkers, not able to find a gap between vehicles, must stop in the middle of the road before they can cross completely.
Though rare, ``dangerous" jaywalkers do exist in real-life, and it is important that safety-critical AV systems have their data.

\begin{table}[t]
    \centering
    \footnotesize
    \begin{subtable}[t]{\columnwidth}
        \centering
        \begin{tabular}{lccc}
            \hline
            \noalign{\smallskip}
            \textbf{\parbox{2.2cm}{Truth Value $\rightarrow$\\ Guessed value $\downarrow$}} & \textbf{Real-life} & \textbf{VR} & \textbf{Synthetic} \\
            \noalign{\smallskip}
            \hline
            \textbf{Real} & 67.0\% & 58.6\% & 10.9\% \\
            \textbf{Fake} & 33.0\% & 41.4\% & 89.1\% \\
            \hline
        \end{tabular}
        \caption{Percentages}
        \label{subtab:3rd_person_without_count}
    \end{subtable}%
    \vskip\baselineskip % Add vertical space between subtables
    \begin{subtable}[t]{\columnwidth}
        \centering
        \begin{tabular}{lccc}
            \hline
            \noalign{\smallskip}
            \textbf{\parbox{2.2cm}{Truth Value $\rightarrow$\\ Guessed value $\downarrow$}} & \textbf{Real-life} & \textbf{VR} & \textbf{Synthetic} \\
            \noalign{\smallskip}
            \hline
            \textbf{Real} & 547 & 482 & 19 \\
            \textbf{Fake} & 269 & 340 & 155 \\
            \hline
            \textbf{Total} & 816 & 822 & 174 \\
            \hline
        \end{tabular}
        \caption{Counts}
        \label{subtab:3rd_person_with_count}
    \end{subtable}
    \caption{Confusion matrix for survey respondent guesses (real / fake) vs. truth category (real-life, VR, synthetic)}
    \label{tab:3rd_person}
\end{table}

\begin{table}[t]
    \centering
    \begin{tabular}{l|p{1.7cm}p{1.7cm}p{1.7cm}}
        \hline
        \noalign{\smallskip}
         \textbf{Pairing} & \textbf{Real-life / VR} & \textbf{Real-life / Synthetic} & \textbf{VR / Synthetic} \\
        \noalign{\smallskip}
        \hline
        \noalign{\smallskip}
         \textbf{\% Correct (N)} & 64.5\% (732) & 89.3\% (84) & 88.9\% (90)\\
        \noalign{\smallskip}
        \hline
    \end{tabular}
    \caption{Percent correct of each pairing. ``Correct" is defined as selecting real as more ``real" when compared with either VR or synthetic, and selecting VR as more ``real" when compared to synthetic.}
    \label{tab:pairings}
\end{table}

\section{Limitations, Discussion, and Future Work}

\subsection{Interview Responses from Part 1 Not Addressed by VR System}
There were some concerns elicited by the participants of the Part 1 structured interview that were not addressed by the evaluated VR pedestrian simulator. 
Some of these concerns form the foundation for future work:
\begin{enumerate}
    \item I1 and I2 suggested that the data from vehicle simulators may be less noisy than real-life data, enlarging the sim-to-real gap.
In the VR CARLA environment, vehicles follow simple policies like the Intelligent Driver Model,
% ~\cite{}
always driving perfectly along predefined paths.
This may not realistically reproduce human driving behavior, which is imperfect, and sometimes deviates from the center of the lane.
In the future, more complex and noisy driving policies can be implemented into the CARLA environment to create more realistic vehicle behavior.
In addition, as described in the concurrent work contributing the VR system~\cite{anon2023}, a real human driver can be hooked into the simulator as an additional vehicle agent. The real human driver will create vehicle trajectories that pedestrians users may deem more natural.

\item I4 pointed out that certain object categories were under-represented in their company's large internal dataset, thus impacting the ability of the AV system to detect them. 
For example, small children, specific types of signage such as construction zone signs, skateboarders, trikes, and bicycles in particular were lacking.
Future work includes inviting children and elderly to collect data with our system, so we can make use of the low-risk benefits of the VR environment to obtain the lacking but much-needed behavioral data of vulnerable road users.
% \item I2 pointed out that she knew of a certain AV company's internal virtual world simulator that offered realistic visuals rendered directly from real-life images. 
% Realistic assets and 3D world scans could be incorporated into the simulator, replacing the game-like CARLA assets. 
% The trend is moving toward building a true underlying 4D representation of the world.  augmentation of real-world data, unlike CARLA.
% In addition, detection of ``generic objects" such as suitcases and luggage carts which did not fall into any particular pre-labelled object category posed a challenge.
\end{enumerate}

\subsection{Free Responses from Part 2 Eliciting System Limitations}
The qualitative responses summarized in Section~\ref{sec:pt2_realism} to the free response question, \textit{What aspects of the system or environment were unrealistic?}, revealed certain limitations of the VR system that can be expanded in future work.
While some concerns are difficult to address with the state of current technologies, such as (1), (2), and (4), other concerns can be resolved with additional engineering of the VR system or revised study design.
To address (3), giving participants a longer time to familiarize themselves with the VR headset may increase their reassurance that they will not run into obstacles in the real world.
To address (5), sound can be easily added to the simulation as it is available in CARLA.
However, we chose not to use it in our study, so that participants could hear us when we verbally instructed participants to where they should walk.
(6) can be addressed by adding into the scene the relevant environmental elements, all of which are available as assets in Unreal Engine.
(7) may be a bit more difficult to address across all environments, but can be addressed for the sidewalk / curb situation by adding physical objects in the real world to match those in the virtual world, and carefully calibrating the virtual-to-real mapping.
% (9) can be addressed
% (10) can be adders
(8) can be addressed by integrating into the system the driving simulator extension, which has already been developed. 
A human driver participant can then use the driving simulator to operate the vehicles in more natural way.
Furthermore, the vehicles currently move based on default CARLA policies; better policies can be integrated into CARLA for more realistic vehicle movement.
(10) can be addressed by adding in additional pedestrians into the simulation, which the system also supports;
each pedestrian would require an additional VR setup, including headset, wireless tracker, and computer. Alternatively, autonomous pedestrians could be added into the simulator. Although these pedestrians would move based on simple policies, these policies are reactive to the user's movements.
To address (11), participants can be given tasks to do rather than specific goals to navigate to,
such as ``Please drop off a letter in the nearest mail box," 
rather than ``Please walk across the road."

With some design modifications and system engineering, nearly all of the concerns raised by the users can be addressed.

\subsection{Part 3 Study Design Choice Limitations}
One limitation of the design of our evaluative study is that there may not be enough information from a simple 2D schematic for a human evaluator to determine if a trajectory is ``real." 
As shown by the results, there is enough information present in the 2D trajectory schematics to distinguish between a real-life and a simple synthetic linear policy, as well as between a VR and a simple synthetic linear policy.
However, the evaluator has no access to information that could be used to make a better decision, such as body pose. 
As future work, a body-pose data collection system can be set up in real-life and used to collect body pose data from real-life jaywalkers. 
The body pose data can be used to create animations for a third-person 3D full-body trajectory comparative evaluation. 

Again, it is important to note that just because a VR-collected trajectory is distinguishable from a real trajectory by a human evaluator, that does not necessarily imply that it is unrealistic or unseen in real-life. 
The VR system specializes in collecting data of uncommon scenarios. This means that even though the VR-collected data may exhibit aggressive behavior, it is behavior that may still be seen in real-life.

% Another limitation is that it is uncertain the degree to which the HTC VIVE system smooths the tracked trajectories during VR data collection. Thus, an estimate had to be used to smooth the real-life trajectories.

% Another limitation is that 

% \section{Conclusion}
% We evaluated the degree to which a VR data collection system can elicit genuine pedestrian responses, confirming that such a system addresses many of the current limitations of data collection methods. Many aspects of the virtual environment of the system felt real to our users. Further, trajectories collected from the VR system appeared much more realistic than synthetically-generated data to third-person evaluators. Vulnerable road users such as children and elderly behave differently than  are Overall, the VR system offers a low-risk method of collecting data of uncommon pedestrian-vehicle interactions, opening the way for  vulnerable road users.

\section{Conclusion}
Safely and ethically collecting pedestrian trajectory data for training reliable autonomous vehicle models is a complex and nuanced problem; but Virtual Reality data collection offers a solution. A VR data collection system fills the need for diverse and comprehensive data while overcoming the difficulty of obtaining such data in the real world. Through semi-structured interviews, a first-person evaluation, and a third-person survey, we provide substantial evidence that VR-based system can evoke a genuine response from pedestrian users, and thus offers an effective data collection solution.

Despite some limitations in the sense of presence within the virtual environment, the data collected still appears to be reasonably similar to that collected in real-life scenarios. This presents a compelling case for VR as an effective tool for gathering pedestrian data, particularly in complex, high-risk, or ethically challenging scenarios that are difficult to replicate in the real world. These findings add a new dimension to our understanding of vehicle-pedestrian data collection, emphasizing the potential for VR technology to play a vital role in shaping the future of safe and reliable autonomous transport.

%% Use plainnat to work nicely with natbib. 

\bibliographystyle{plainnat}
\bibliography{references}

\end{document}